\documentclass[jkps,twoside,twocolumn,showpacs,showkeys,floatfix]{revtex4}
\usepackage{amssymb,amsmath}
\usepackage[pdftex]{graphicx}

\begin{document}

\title{Effects of Janus Oscillators in the Kuramoto Model  with Positive and Negative Couplings }

\author{Jungzae \surname{Choi}}
\affiliation{Department of Physics and Department of Chemical Engineering, Keimyung University, Daegu 42601, Korea}
\author{MooYoung \surname{Choi}}
\affiliation{Department of Physics and Astronomy and Center for Theoretical Physics, Seoul National University, Seoul 08826, Korea}
\author{Byung-Gook \surname{Yoon}}
\thanks{E-mail: bgyoon@ulsan.ac.kr}
\affiliation{Department of Physics, University of Ulsan, Ulsan 44610, Korea}

\begin{abstract}
We study the effects of Janus oscillators in a system of phase oscillators in which the coupling constants take both positive and negative values. Janus oscillators may also form a cluster when the other ones are ordered, and we calculate numerically the traveling
speed of three clusters emerging in the system and average separations between them, as well as the order parameters for three groups of oscillators, as the coupling constants and the fractions of  positive and Janus oscillators are varied.
An expression explaining the dependence of the traveling speed on these parameters is obtained and observed to fit well the numerical data. With the help of this, we describe how Janus oscillators affect the traveling of the clusters in the system.

\end{abstract}

\maketitle

\section{Introduction}

A system of coupled phase oscillators is a typical model
exhibiting collective synchronization behavior, which is revealed as a cluster formation on a phase circle.
The model was first introduced by Winfree~\cite{ref:Winfree}, and later refined by
Kuramoto~\cite{ref:Kuramoto0,ref:Kuramoto}. Since these works, many variations of the Kuramoto
model have been published~\cite{ref:extension}. One interesting extension of the model in view of cluster formation
is  incorporating both repulsive and attractive couplings~\cite{ref:hs1,ref:hs2}.

Each oscillator is identified either as a negative (repulsive) or as a positive (attractive) one;
accordingly, two clusters, which are separated on a phase circle, may form.
When this phase separation is less than $\pi$ radians, the two clusters travel
on the phase circle, maintaining an average separation angle. Recently, the dynamics of the traveling
state in this model, as well as in a variant, has been studied in terms of the phase speed
of clusters~\cite{ref:choij14,ref:choij18}.

Systems suggested as a possible application of this model include the sociophysical models of
opinion formation~\cite{ref:opinion}, for which the positive/negative coupling corresponds to
conformists/contrarians~\cite{ref:hs1}.
Thus, in this context, the incorporation of 
intermediate group in the model would be interesting. One possibility for this is a group of constituents with two-faced characters, in which
an oscillator interacts attractively/repulsively with the positive/negative ones and hence the name ``Janus" oscillator~\cite{ref:janus}.
In this work, we study numerically an oscillator model in which Janus oscillators are also present.
We observe that a cluster of Janus oscillators may form and travel, in addition to the two clusters of positive and negative oscillators.
We calculate the traveling speed of the three clusters emerging in the system, the average separations between them, and the order parameters  for the three groups of oscillators as the coupling constants and the compositions of the three types of oscillators are varied.
An expression explaining the dependence of the traveling speed on these parameters is obtained and observed to fit well the numerical data. With the help of this, we describe how Janus oscillators affect  the traveling speed of the clusters in the system.

This paper consists of four sections: In Section II, the oscillator model and its dynamics are described. Section III presents numerical results, together with
phenomenological interpretations of the effects of Janus oscillators.
Finally, a brief summary is given in Section IV.

\section{Model and Numerical calculation}

We consider a system of $N$ oscillators, the $i$th of which has intrinsic frequency $\omega_i$. The oscillator is described by
its phase and is coupled globally to other oscillators.
The dynamics of such a coupled oscillator system is governed
by the set of equations of motion for the phase $\phi_i$
of the $i$th oscillator ($i=1,...,N$):
\begin{equation} \label{model}
 \dot{\phi_i} =\omega_i -  \frac{1}{N}\sum_{j=1}^N K_i \sin(\phi_i-\phi_j),
\end{equation}
where the intrinsic frequencies are assumed to be symmetrically distributed according to the
Lorentzian distribution
$g(\omega)= (\gamma_{\omega}/\pi) (\omega^2 + \gamma_{\omega}^2 )^{-1}$.
The other term on the right-hand side represents
sinusoidal interactions with other oscillators, where the coupling constant $K_i$ takes
a positive or negative value. A new feature in this work is that a fraction $q$ of oscillators
are of Janus character, which means that the coupling constant for such a Janus oscillator takes
 a positive/negative value when it interacts with a positive/negative oscillator. Thus, for this oscillator,
 the coupling constant $K_J$ takes a value $K_{Jp}/K_{Jn}$ when it couples with a positive/negative oscillator.
Further, we take $K_J=0$ between Janus oscillators, which means Janus oscillators do not interact with one another.
Specifically, the coupling is taken from the distribution
$\Gamma(K)=(1{-}q)p\delta(K{-}K_{p})+(1{-}q)(1{-}p)\delta(K{-}K_{n})+q\delta(K{-}K_{J})$, where
$K_{p/n}$ is the positive/negative coupling constant ($K_{p} >0$ and $K_{n} <0$)
and $p$ is the fraction of normal oscillators having a positive coupling constant.

In order to measure the synchronization of the system, we
introduce the complex order parameter
\begin{equation} \label{deforder}
  \Psi \equiv \frac{1}{N} \sum_{j=1}^N e^{i \phi_j}
       = \Delta e^{i\theta} ,
\end{equation}
which characterizes the synchronization of the oscillators with the magnitude $\Delta$ and the average phase $\theta$. The order parameter defined in
Eq.~(\ref{deforder}) allows us to reduce Eq.~(\ref{model}) to a single decoupled equation:
\begin{equation} \label{eqn}
 \dot{\phi_i} = \omega_i - K_i \Delta \sin(\phi_i -\theta).
\end{equation}
To investigate the behavior of the system governed by Eq. (\ref{eqn}), we resort mainly to numerical methods.
Using the second-order Runge-Kutta algorithm, we integrate
Eq.~(\ref{eqn}) with the time step $\Delta t=0.01$ for the system size
$N{=}2000$. Initially ($t{=}0$), $\phi_i$'s are
randomly distributed between $0$ and $2\pi$ for all $i$. We fix the positive coupling constant to be $K_{p} =1$
and set the ratio $K_{Jn} /K_{Jp} $ equal to $K_n /K_p$ such that $K_{Jn}=K_{Jp} K_n $ throughout this work.

After the initial transient behavior, the system becomes stationary, and we obtain a time series of 
order parameter information on the time evolution of the phase distribution.
In order to understand the clustering behavior,
we define the order parameter of the positive/negative oscillators $\Psi_{p/n} = \Delta_{p/n} e^{i\theta_{p/n}}$, as well as that of the Janus oscillators $\Psi_{J} = \Delta_{J} e^{i\theta_{J}}$, similarly to Eq.~(\ref{deforder}). These parameters and the phase separations $\delta_{pn}$, $\delta_{Jp}$ and $\delta_{Jn}$ between the phases of the three kinds of oscillators are also calculated in each run.
Namely, for example, $\delta_{Jp}\equiv |\theta_J - \theta_p |$ stands for the angular distance between the two phases $\theta_J$ and $\theta_p$ for Janus and positive oscillators, respectively, and likewise for $\delta_{pn}$ and $\delta_{Jn}$.

The average values of these, as well as the traveling speed $w$, are obtained in the following way:
At each time, we calculate these parameters and the average value of the phase velocities $w_i = \dot{\phi}_i$ over the oscillators.
Then, we get the time average over $10^{4}$ time steps and take the absolute value to obtain the traveling speed.
The parameters $\Delta$'s and $\theta$'s do not vary much in time, and we usually take the values after the time evolution.
Finally, we take the averages over 30 initial configurations to obtain the (average) speed $w$.
We have also considered 100 initial configurations to average over and varied the size $N$ of the system up to 6000 to find that the overall behaviors are unaltered.
We also examine populations in given regions of phase space.
Specifically, we divide one cycle of the phase angle into
72 different ranges of equal width, and we obtain
the number $n(\phi)$ of oscillators belonging to a range centered at $\phi$.

The traveling speed $w$ obtained numerically can be explained essentially in
the same way as in Refs.~7 and 8.
Suppose that $N p(1{-}q) \Delta_p$ oscillators with the phase $\theta_p$ are in the positive cluster,
$N(1{-}p)(1{-}q) \Delta_n$ oscillators with the phase $\theta_n$  are in the negative cluster, and
$Nq \Delta_J$ oscillators with the phase $\theta_J$ are in the third (Janus) cluster.
Assuming that the remaining oscillators are desynchronized and have no net effects on $w$,
one can derive straightforwardly, from Eq.~(\ref{model})
\begin{align}
 \label{wformula}
     w = & \Delta_n \Delta_p (K_p - K_n ) p(1-p) (1-q)^2 \sin \delta_{pn}           \nonumber \\
           &+\Delta_J \Delta_p (K_{Jp}- K_p ) pq(1-q) \sin \delta_{Jp}           \nonumber \\
            &+\Delta_J \Delta_n (K_{Jn}- K_n )(1-p)q(1-q) \sin \delta_{Jn}.
\end{align}
Note that only the first term in Eq. (\ref{wformula}) remains when $K_{p}=K_{Jp}$ and  $K_{n}= K_{Jn}$,
the condition under which the data in the last three figures of this work have been calculated
 (as well as for a system without Janus oscillators).
This equation, when the width $\gamma_{\omega}$ of the frequency distribution is small,
has been found to give a good description of the data for the traveling speed in this work.
If $\gamma_{\omega}$ becomes too large, the system becomes disordered, and no clusters are formed.

\section{Results and discussion}

\begin{figure}
\includegraphics[width=8.5cm]{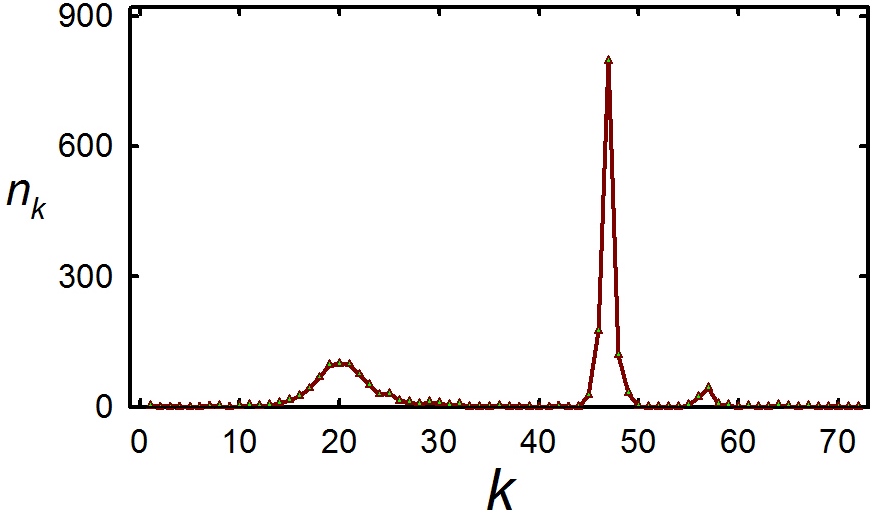}
\caption{(color online)  Number of oscillators $n(\phi)$ belonging to a domain of width $\pi/36$ centered at $\phi$ (modulo $2\pi$)
in a system of $N=2000$ oscillators with $q=0.04$, $K_{Jp}=0.5$, $K_n={-}0.5$, $p=0.6$ and
$\gamma_{\omega}=0.01$. The distribution has been obtained from a steady traveling state. The line is merely a guide for the eyes.  Note that
$K_{Jn}=K_{Jp}K_n$ (setting $K_p =1$) throughout this work (See text).
}
\label{fig:distrib}
\end{figure}

Before explaining the traveling of the three clusters, we describe how the traveling occurs in a system of positive and negative oscillators without Janus ones.
When the positive coupling is sufficiently large and the positive cluster is formed, negative oscillators move away
from the positive cluster. This may result in the formation of a negative cluster if 
the repulsive interactions between negative oscillators can be overcome.
If this repulsion between two clusters is too strong, the separation angle becomes $\pi$ radians and
no traveling occurs, which is easily understood in view of Eq.~(\ref{wformula}) with $q=0$ or Eq.~(\ref{model}).
Thus, we find that the clusters travel only if the separation is less than  $\pi$.

When Janus oscillators are present together with positive and negative clusters,
they move away from the negative oscillators farther than they move away from the positive ones
because a Janus oscillator does not attract negative ones like a positive oscillator does.
Consider a system in which  the repulsive interaction, as well as the Janus coupling,
is strong enough so that a Janus cluster is formed at the farthest angle $\pi$ from the negative cluster.
Then, the positive cluster will be located at an angle between phases of the negative and the Janus ones;
otherwise, the system would become unstable. The distance between the Janus and the positive clusters $\delta_{Jp}$
depends on the strength of the overall repulsive interaction due to the negative cluster.
Note that this overall interaction depends on the coupling constants and the phase separations, as well as the populations of the clusters.
The separation $\delta_{pn}$ is important in determining the traveling speed of this system. To see this, we may examine Eq.~(\ref{wformula}):
Since $\delta_{Jn}$ is close to, but slightly less than, $\pi$, 
the third term contributing to $w$ is nearly zero. Moreover, we have
$\sin \delta_{pn}{\approx}\sin\delta_{Jp}$ because
$\delta_{Jp}$ is approximately equal to $\pi {-} \delta_{pn}$.
Thus, the traveling speed is roughly proportional to $\sin \delta_{pn}$.

Of course, we should resort to the full form of Eq.~(\ref{wformula}) for the other systems in which $\delta_{Jn}$ is
somewhat less than $\pi$ radians and/or $\delta_{Jp}$ is larger than, say, one radian.
Two cases are met: either the sum of the three separation angles is $2\pi$ radians
or $\delta_{Jn}=\delta_{pn}+\delta_{Jp}(<\pi)$.
We find that the former corresponds to a system in which the overall repulsive interaction due to negative oscillators is very weak, and the population of Janus cluster is very small.

\begin{figure}
\includegraphics[width=8cm]{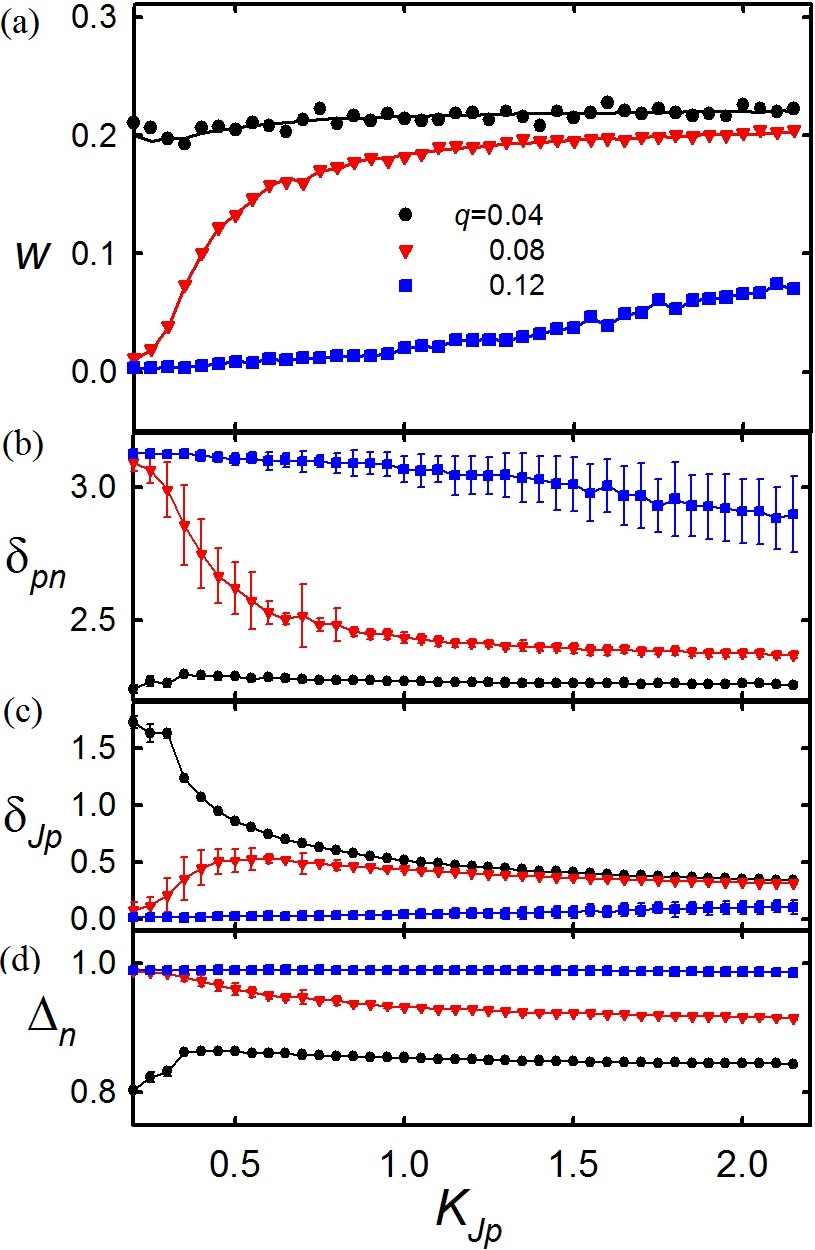}
\caption{(color online) (a) Average traveling speed $w$, phase separations (b) $\delta_{pn}$ between positive and negative clusters and (c) $\delta_{Jp}$ between positive and Janus clusters, and (d) order parameter $\Delta_n$ versus coupling constant $K_{Jp}$ in a system of $N{=}2000$ oscillators with $p=0.6$, $K_n ={-}0.5$ and
$\gamma_{\omega}=0.01$ for three values of the fraction $q$ of Janus oscillators as shown in the legend.
The lines in (a) are plots of Eq.~(\ref{wformula}) whereas the lines in (b) to (d)
are merely guides for the eyes.  Error bars represent standard deviations.
}
\label{fig:vsKjp}
\end{figure}

\begin{figure}
\includegraphics[width=8.5cm]{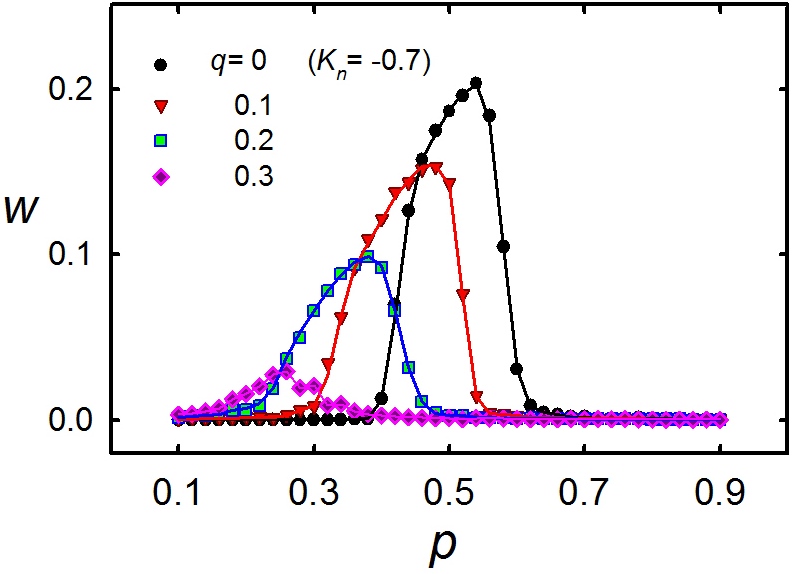}
\caption{(color online ) Average traveling speed $w$ versus the fraction $p$
 in a system of $N{=}2000$ oscillators with $\gamma_{\omega}=0.01$, $K_n={-}0.7$ and $K_{Jp}=1$ for four values of $q$ as shown in the legend.
Lines are plots of Eq.~(\ref{wformula}) whereas symbols represent the data points obtained numerically.
}
\label{fig:wvsp}
\end{figure}

\begin{figure}
\includegraphics[width=8cm]{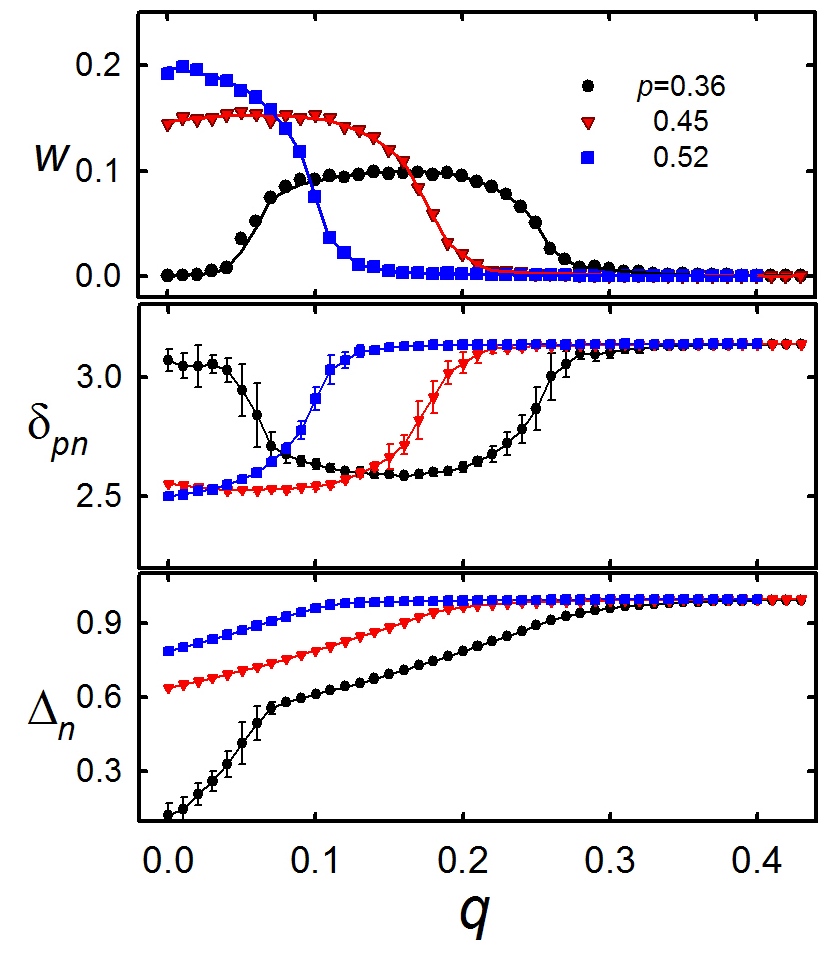}
\caption{(color online) (a) Average traveling speed $w$, (b) phase separation $\delta_{pn}$ between positive and negative clusters, and (c) order parameter $\Delta_n$ versus $q$ in a system of $N{=}2000$ oscillators with $K_n ={-}0.7$, $K_{Jp}=1$, and $\gamma_{\omega}=0.01$ for three values of $p$ as shown in the legend. The lines in (a) plot Eq.~(\ref{wformula}) whereas the lines in (b) and (c)
are merely guides for the eyes.  Error bars represent standard deviations.
}
\label{fig:vsq}
\end{figure}

\begin{figure}
\includegraphics[width=7cm]{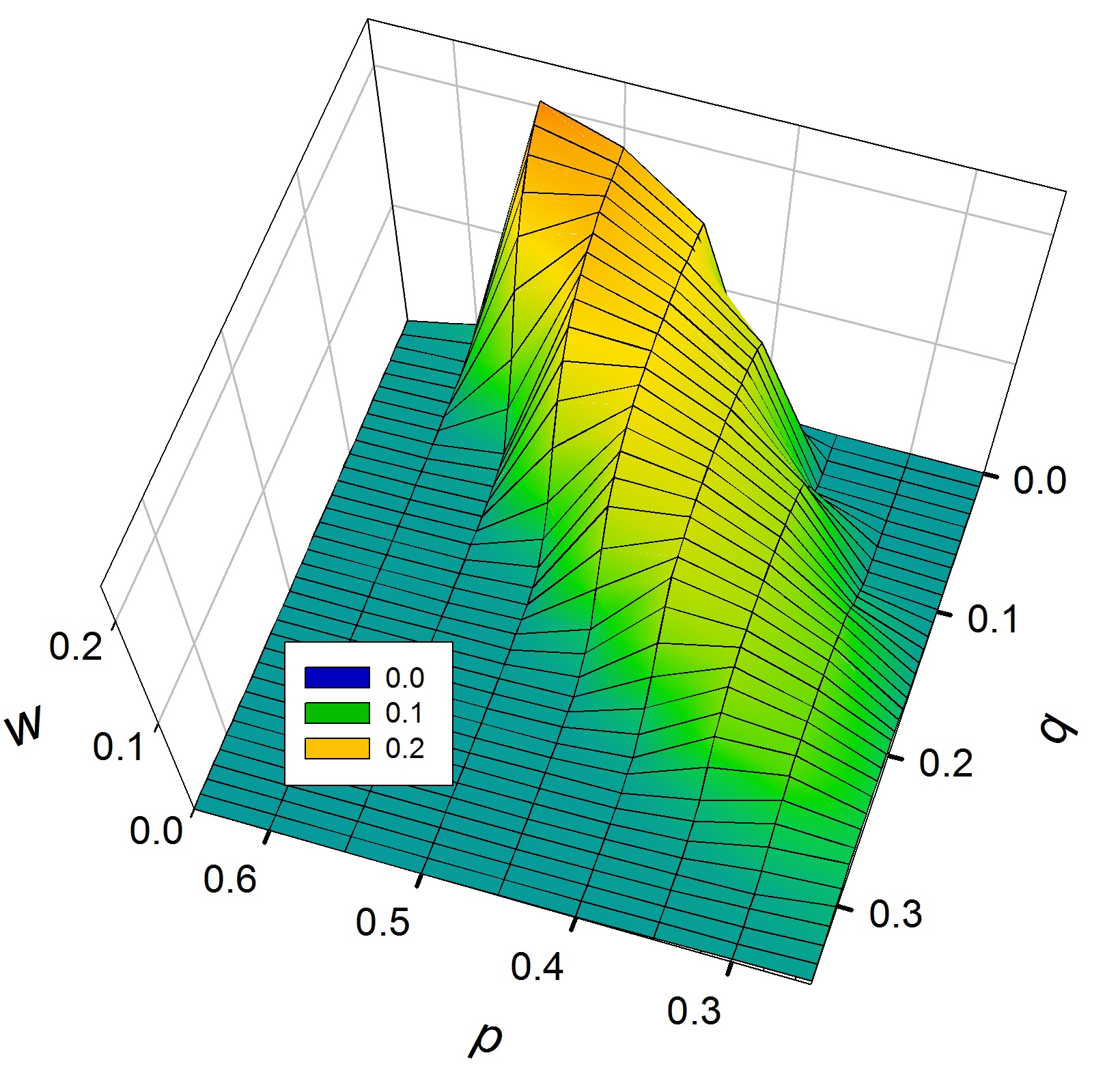}
\caption{(color online)  Average traveling speed $w$ versus $p$ and $q$ in a system of $N{=}2000$ oscillators with $K_n ={-}0.7$, $K_{Jp}=1$, and $\gamma_{\omega}=0.01$.
}
\label{fig:wvspq}
\end{figure}
With these ideas in mind, we now present our numerical results.
Figure~\ref{fig:distrib}, which has been obtained from a system having $\delta_{Jn}{=}\pi$,
shows the number of oscillators $n(\phi)$ belonging to a domain of width $\pi/36$ centered at $\phi$ (modulo $2\pi$)
 in a system of $N{=}2000$ oscillators with $q=0.04$, $K_{Jp}=0.5$, $K_n={-}0.5$, $p=0.6$ and $\gamma_{\omega}=0.01$.
Three clusters are clearly observed: negative, positive and Janus clusters from the left.
The value of the separation $\delta_{Jp}$ is about $0.9$ in this figure. Here, one may
note that the clustering of mutually noninteracting members is less surprising than that of mutually repelling members.

We then show how the system behaves as the Janus coupling strength is varied.
Figure~\ref{fig:vsKjp} displays (a) the average traveling speed $w$, (b) the phase separation $\delta_{pn}$ between positive and negative clusters, (c) the phase separation $\delta_{Jp}$ between positive and Janus clusters, and (d) the order parameter $\Delta_n$ plotted versus the Janus coupling constant $K_{Jp}$ in a system of oscillators
with $p=0.6$, $K_n ={-}0.5$, and $\gamma_{\omega}=0.01$ for three values of the fraction $q$ of Janus oscillators as shown in the legend. At the lowest value of $K_{Jp}=0.2$ in the figure, the system with $q=0.04$ is in the traveling state, as revealed in (a) and (b), while $w$ is very small for the other larger values of $q$, leading to a standing (non-traveling) state, which we call a $\pi$-state after Ref.~5.
Without Janus oscillators, the system will be in the traveling state. In like manner,
a small number of Janus oscillators do not have appreciable effects on the separation $\delta_{pn}$
between positive and negative oscillators. As the value of $q$ increases, $\delta_{pn}$ increases
toward $\pi$. This can be explained as follows: Once positive oscillators are
ordered, the positive cluster will attract Janus oscillators, forming a mixed cluster of positive and Janus oscillators.
From the viewpoint of the negative cluster, the size of the cluster from which it should run away has increased noticeably.
The results are the increased order parameter $\Delta_{n}=1$, as is shown in (d), and
the increased separation $\delta_{pn}$ or the $\pi$ state.
Here, the presence itself of Janus oscillators causes this behavior, while their couplings with the other clusters are negligible.

As the Janus coupling constant $K_{Jp}$ is increased, dramatic changes
arise in the data for $q=0.08$; the positive cluster moves away from the Janus cluster, and the traveling speed increases as a result.  This motion is not easy to explain thoroughly because it results from intriguing attractive
and repulsive interactions among the three clusters. Roughly speaking, the configuration such that the positive cluster
is separated from the Janus one is usually more favorable in the presence of Janus coupling,
except when the overall repulsion due to the negative cluster is very strong.
Otherwise, due to the Janus coupling, this oscillator may move in both directions at the $\pi$ position if the negative oscillators are
not almost completely ordered. Thus, the positive oscillators move slightly even for a small value of $K_{Jp}$.
The slight increase in $w$ with increasing $K_{Jp}$ in the data for $q=0.12$ can be explained similarly:
As the coupling is increased, the separation $\delta_{pn}$
decreases, but not as much as it does for the data for $q=0.08$ because $\Delta_n{\approx}1$ and the overall effect of the repulsive interaction is very strong.

Now, we move to an issue about the domain of the parameter $p$ in which traveling occurs.
Figure~\ref{fig:wvsp} shows the average traveling speed $w$ versus the fraction $p$
 in a system of $N{=}2000$ oscillators with $\gamma_{\omega}=0.01$, $K_n ={-}0.7$, and $K_{Jp}=1$ for four values of $q$ as shown in the legend.
We found that the domain of the traveling state moves towards the origin of the $p$ axis and broadens as $q$ is increased, although the peak speed is decreased.
An argument similar to that in the last paragraph can be put forward on
how a $\pi$-state becomes a traveling state in the presence of Janus oscillators at smaller values of $p$ and vice versa at larger values of $p$.

In doing this, an inspection of Fig.~\ref{fig:vsq}, which presents (a) the average traveling speed $w$, (b) the phase separation $\delta_{pn}$ between positive and negative clusters, and (c) the order parameter $\Delta_n$ versus the fraction $q$ in a system of $N{=}2000$ oscillators with $K_n ={-}0.7$, $K_{Jp}=1$, and $\gamma_{\omega}=0.01$ for three values of $p$ as shown in the legend, would be helpful.
When the fraction of positive oscillators is lower ($p=0.36$) in the absence of
Janus oscillators ($q=0$), the system is in the $\pi$-state. In this case, the order parameter of the negative cluster is very small, as observed in Fig.~\ref{fig:vsq}(c), because positive oscillators are not enough to cause negative oscillators to be well ordered.
If Janus oscillators are introduced in the system with these parameters, they are usually clustered in a way that $\delta_{Jn}$ is close to $\pi$ radians. If the fraction $q$ is raised toward $0.1$, the number of oscillators interacting with
negative oscillators becomes so large as to have $\Delta_n$ grow remarkably.
Because the negative coupling is not so strong and the order parameter $\Delta_n$ is not that large, the separation $\delta_{pn}$ reduces below $\pi$ (and clusters begin to travel). When the fraction $p$ is high ($p=0.52$) with $q=0$, on the other hand, the traveling speed is large, and the phase separation is given by $\delta_{pn}=2.5$. As the fraction is increased to $q=0.15$, the clusters cease to travel. In this case, the Janus cluster causes the order parameter $\Delta_n$ to grow to a value larger than $0.9$, which, in turn, causes the separation angle $\delta_{pn}$ to increase, resulting in a $\pi$-state.
Finally, the main result is summarized in Fig.~\ref{fig:wvspq} which presents the average traveling speed $w$ versus $p$ and $q$ in a system of $N{=}2000$ oscillators with $K_n ={-}0.7$, $K_{Jp}=1$, and $\gamma_{\omega}=0.01$. Serving as a phase diagram for the traveling (or the $ \pi $) state in $(p,q)$ space, it is helpful in understanding the discussion on the
behavior.

\section{Summary}

We have considered a system of phase oscillators in which the coupling constants take both positive and negative values and have probed the effects of Janus oscillators.
Janus oscillators have also been found to form a cluster when the positive and the negative
oscillators are ordered. The resulting three clusters are traveling when the phase separation between positive and negative clusters is less than $\pi$ radians.
We have obtained an expression explaining the dependence of the traveling speed on these parameters and observed that it fit well the numerical data. With the help of this, we have described how Janus oscillators affect the traveling of the clusters in the system. Depending on the parameters of the system, such Janus oscillators may facilitate or hinder traveling.

\section*{Acknowledgment}
This work was supported in part by the 2019 Research Fund of the University of Ulsan.

\end{document}